\documentclass{epl}

\usepackage{amsmath}

\title{Four-path interference and uncertainty principle in photodetachment microscopy}
\shorttitle{Four-path interference and uncertainty principle etc.}
\author{T. Kramer\thanks{E-mail: \email{tkramer@ph.tum.de}} \and C. Bracher \and M. Kleber}
\shortauthor{T. Kramer \etal}
\institute{
Physik-Department T30, Technische Universit\"at M\"unchen\\ James-Franck-Stra{\ss}e, 85747~Garching, Germany}

\begin{document}

\maketitle

\begin{abstract}
We study the quantal motion of electrons emitted by a pointlike monochromatic isotropic source into parallel uniform electric and magnetic fields. The two-path interference pattern in the emerging electron wave due to the electric force is modified by the magnetic lens effect which periodically focuses the beam into narrow filaments along the symmetry axis. There, four classical paths interfere. With increasing electron energy, the current distribution changes from a quantum regime governed by the uncertainty principle, to an intricate spatial pattern that yields to a semiclassical analysis.
\end{abstract}

\section{Introduction} 
\begin{figure}
\onefigure{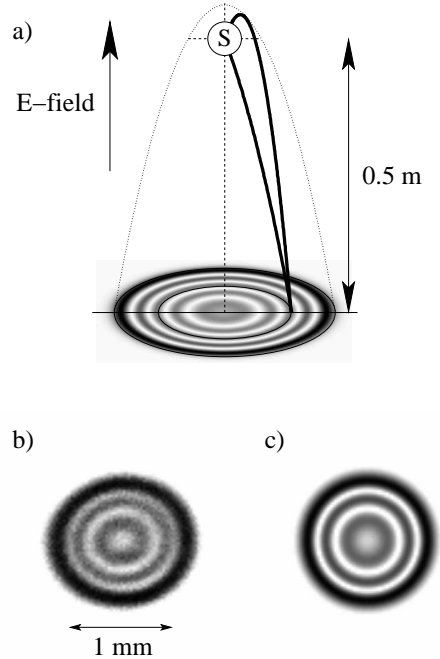}
\caption{Photodetachment in a uniform electric field environment. (a) Schematic view of the photodetachment microscope used by Blondel et~al.\ \protect\cite{BlondelI,BlondelII,BlondelIII}. Electrons released with constant energy at the source region (labeled $S$) are accelerated towards the detector, (b) recorded image of the current density for O$^-$ photodetachment, (c) theoretical prediction as obtained from Eq.~(\protect\ref{eq:CurrentDensity}), dark rings correspond to higher current density, parameters: $B=0$, $E=100.5$~$\mu$eV, $F=423$~eV/m, $z=-0.514$~m.}
\label{setup_exp}
\end{figure}

Two-path interference along classical trajectories has a long-standing tradition as a textbook showpiece of quantum mechanics \cite{FeynmanIII}.  A fascinating realization of two-path interference on a macroscopic scale has recently been achieved in near threshold photodetachment microscopy \cite{BlondelI,BlondelII,BlondelIII}.  Here, electrons are released from negative ions by irradiation with a laser beam in the presence of a uniform external electric field which subsequently governs their motion.  Blondel et~al.\ \cite{BlondelI,BlondelII,BlondelIII} recorded field-induced interference fringes in configuration space for O$^-$ and several other ionic species.  In their experiment, the motion of the electrons can be considered as the quantum analogue of throwing a classical particle at constant energy in a uniform gravitational field. Within the classically allowed ``shot-put''-range two distinct trajectories will link the electron source with a given destination, causing ``double slit'' interference in the quantal case. 
Using detectors with high spatial resolution, images of the resulting fringes were obtained by Blondel et~al.\ \cite{BlondelII,BlondelIII} at large distances ($0.5$~m) from the electron-emitting negative ions, there extending to the millimeter scale (Fig.~\ref{setup_exp}). In this way, their ``photodetachment microscope'' demonstrated the nodes and antinodes of the electronic wave function, allowing precise determination of the electron affinity\ \cite{BlondelIII}. In order to advance the understanding of the imaging mechanism of their device, this letter serves to point out the intricate spatial properties of the photoelectronic current distribution in parallel electric and magnetic fields.  (The integrated photocurrent, i.~e.\ the total photodetachment cross section in parallel fields has been addressed theoretically by several authors \cite{Du,Fabrikant}.  Their results are implicitly contained in the following developments.)

For photon energies close to the electron affinity of the ion, the photodetachment process becomes insensitive to details of the initial and final atomic states involved, and the shape of the electronic scattering wave in the external field environment is determined by its parity and the electron excess energy $E$ alone \cite{Farley}.  In a formal description, we may thus replace the photon-ion interaction as an electron-generating mechanism by a fixed pointlike ``source'' that emits electrons with proper angular characteristics \cite{TunnelQBM}.  On grounds of simplicity, in this Letter we only consider the simplest case of isotropic emission that applies to Blondel's choice of O$^-$ as ionic species. 

Fig.~\ref{setup_exp} illustrates that for a purely electric field, Blondel's results \cite{BlondelI,BlondelII,BlondelIII} and the predictions from the source model are in excellent accordance. It should be pointed out that the experimental setup was shielded against magnetic fields. Indeed, fields as weak as the earth magnetic field ($B\sim10^{-5}$~T) will already change the interference pattern appreciably. For parallel $\mathbf E$-- and $\mathbf B$--fields, the Lorentz force will cause circular cyclotron motion in the plane perpendicular to the fields, while uniform acceleration takes place in the electric field direction.  Classically, all electrons emitted by a point source at the origin ${\mathbf r}'={\mathbf o}$ will return to the symmetry axis of the system after times of flight $T_k = k\pi/\omega_L$ ($k=1,2,\ldots$), where ${\bm{\omega}}_L=e{\mathbf B}/(2m)$ is the Larmor frequency of the electrons, and $e/m$ their charge to mass ratio.  Hence, in a temporal scheme, the lateral electron distribution is periodically refocused into a sharply defined ``resonance.'' Since the distance of an electron from the source is largely determined by its time of flight, these resonances translate into a series of narrow spatial constrictions of the electronic current along the symmetry axis of the system located around ${\mathbf r}_k = {\mathbf F}T_k^2/(2m)$, where ${\mathbf F} = e{\mathbf E}$ denotes the electric force:  The external potential acts as an ``electromagnetic lens.''  Because the analogy between the temporal and the spatial picture is not perfect, however, the projection of the electron source is distorted into a filament structure whose dimensions grow with increasing electron energy.  In this Letter, our main attention is directed towards these resonance regions.

\begin{figure}
\onefigure{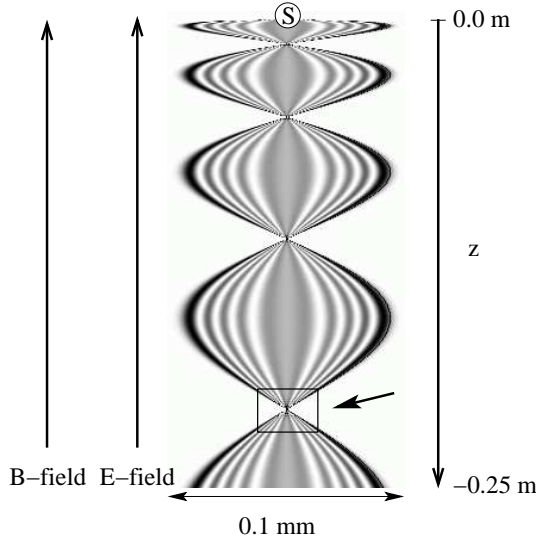}
\caption{Plot of the current density distribution obtained from Eq.~(\protect\ref{eq:CurrentDensity}) for parallel fields. The fourth resonance (arrow) is plotted in detail in Fig.~\protect\ref{res_detail}. Parameters: $E=60.8$~$\mu$eV, $F=116$~eV/m, $B=0.001$~T. There is rotational symmetry about the $z$-axis.}
\label{res_overview}
\end{figure}

A thorough classical analysis of the problem shows that in a uniform electric field environment, within a range of paraboloid shape always two particle trajectories of energy $E$ exist that connect the source ${\mathbf r}' = {\mathbf o}$ with a given destination $\mathbf r$ \cite{TunnelQBM}.  An additional parallel magnetic field will split this continuous sector of two-path degeneracy into a series of distinct regions with finite spatial extension, as illustrated in Fig.~\ref{res_overview}.  For weak fields, adjacent structures barely overlap, forming the ``bottlenecks'' that characterize the current resonances.  Consequently, here the turning surfaces of the intersecting two-path regions delineate a sector where the number of classical trajectories joining source and destination is doubled to four.  Fig.~\ref{res_detail} displays a section of the resonance area with its intertwining caustic surface structure.

In near-threshold emission, the resolution of this electromagnetic microscope is limited by diffraction, i.e., by the uncertainty constraint $\Delta p_{\perp}\,\Delta r_{\perp} \geq \hbar$ \cite{foot_uncertainty}. In this regime the resonances are characterized by a Gaussian current profile, whose typical diameter ($\Delta r_{\perp}\approx 100$ nm) much exceeds the size of the emitting ions.

\section{Source theory of photodetachment} In our model, we introduce a stationary source term $\sigma({\mathbf r})$ \cite{Thaler,STM} to the stationary Schr\"odinger equation:
\begin{equation}\label{eq:psi_source}
\left( E-{\hat H}_{\text{qbm}} \right) \psi({\mathbf r};E)=\sigma({\mathbf r}).
\end{equation}
Here, ${\hat H}_{\text{qbm}}$ is the Hamiltonian for quantum ballistic motion in a uniform electric and magnetic field
\begin{equation}\label{eq:HQBM}
{\hat H}_{\text{qbm}}=\frac{{\mathbf p}^2}{2m}
   -{\bm{\omega}}_{L}\cdot{\mathbf L}
   +\frac{m}{2}{\left[\bm{\omega}_L\times{\mathbf r}\right]}^2
   -{\mathbf r}\cdot{\mathbf F},
\end{equation}
where ${\mathbf L}$ denotes the orbital angular momentum \cite{foot_spin}. By choosing $\sigma({\mathbf r})=\delta({\mathbf r})$ we confine ourselves to a point-like isotropic $s$--wave source of unit strength at the origin ${\mathbf o}$. In this case, Eq.~(\ref{eq:psi_source}) is formally solved by the appropriate Green function $G({\mathbf r},{\mathbf o};E)$:
\begin{equation}\label{eq:G_source}
\left( E-{\hat H}_{\text{qbm}} \right) G({\mathbf r},{\mathbf o};E)=\delta({\mathbf r}),
\end{equation}
and the associated current density reads
\begin{equation}\label{eq:CurrentDensity}
{\mathbf j}({\mathbf r})=
\frac{\hbar}{m}\text{Im}\{ \overline{G({\mathbf r},{\mathbf o};E)}\,\nabla G({\mathbf r},{\mathbf o};E)\}
-(\bm{\omega}_{L}\times{\mathbf r}){|G({\mathbf r},{\mathbf o};E)|}^2.
\end{equation}
The outgoing-wave boundary condition is satisfied only by the retarded Green function $G_{\text{ret}}({\mathbf r},{\mathbf o};E)$. Since ${\hat H}_{\text{qbm}}$ is quadratic in the position and momentum operators, we can use the propagator representation
\begin{equation}\label{eq:propagator}
G({\mathbf r},{\mathbf o};E) = \frac1{i\hbar} \int_0^\infty dt\, a(t)\, \exp\left\{\frac i\hbar\left[ S_{\text{cl}}({\mathbf r},t;{\mathbf o},0) + Et \right]\right\},
\end{equation}
with $S_{\text{cl}}({\mathbf r},t;{\mathbf o},0)$ being the classical action and $a(t)$ a factor independent of ${\mathbf r}$ \cite{FeynmanHibbs}. For uniform electric and magnetic fields aligned to the $z$--axis we obtain
\begin{eqnarray*}
S_{\text{cl}}({\mathbf r},t;{\mathbf o},0)&=&\frac{m\omega_L}{2} \rho^2 \cot\left(\omega_{L}t\right)+\frac{m}{2t}{z}^2
-\frac{Ft}{2}z
-\frac{F^2 t^3}{24 m},\\
a(t) &=& e^{-3i\pi/4}\frac{m\omega_L}{2\pi\hbar\sin(\omega_L t)} \sqrt{\frac{m}{2\pi\hbar t}},
\end{eqnarray*}
with $\rho^2=x^2+y^2$. The right hand side of Eq.~(\ref{eq:propagator}) can be transformed into a sum (see also  \cite{Fabrikant}, Eq.~(11)):
\begin{multline}\label{eq:GreenQM}
G({\mathbf r},{\mathbf o};E)=
\frac{\hbar\omega_{L}}{F\beta}\frac{m^2}{\hbar^4} \exp\left(-\frac{m\omega_{L}}{2\hbar}\rho^2\right)
   \sum_{n=0}^\infty L_n^{(0)}\left(\frac{m\omega_{L}}{\hbar}\rho^2\right) \\
\times\text{Ci}\left\{2\beta\left[Fz-E+\hbar\omega_{L}(2n+1)\right]\right\} \text{Ai}\left\{2\beta\left[  -E+\hbar\omega_{L}(2n+1)\right]\right\}.
\end{multline}
Here, $L_n^{(\alpha)}(x)$ denotes the Laguerre polynomial \cite{Abs}, and $\text{Ci}(x)=\text{Bi}(x)+i\text{Ai}(x)$ is a linear combination of Airy functions. The parameter $\beta = \left[m/(4\hbar^2F^2)\right]^{1/3}$ has the dimension of an inverse energy \cite{foot_AiCi}. Fig.~\ref{res_overview} shows a vertical section of the ensuing current density distribution. Overall, the ring pattern obtained by rotating the current density around the $z$--axis is similar to the one in Fig.~\ref{setup_exp}, but its diameter undergoes repeated oscillations.

\section{Semiclassical approximation} 
\begin{figure}
\onefigure{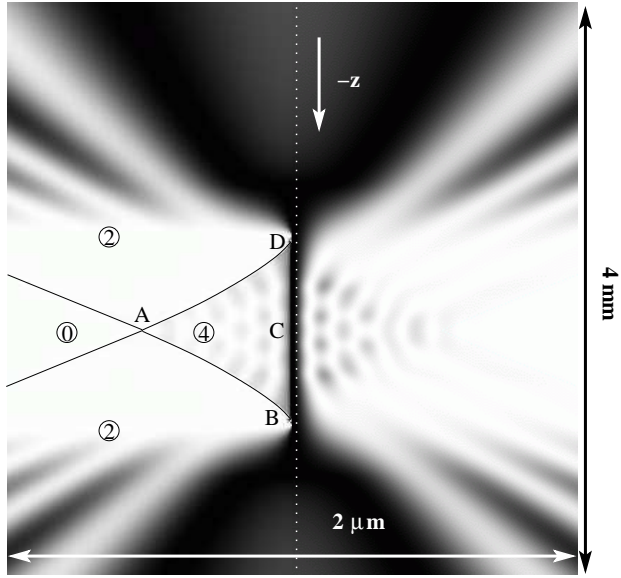}
\caption{Anatomy of the fourth resonance in Fig.~\protect\ref{res_overview}. Circled numbers: Number of classically allowed trajectories. Left hand side: Semiclassical approximation. Right hand side: Quantum solution. Scale: height $4$~mm, width $2$~$\mu$m. Same parameters as in Fig.~\protect\ref{res_overview}.}
\label{res_detail}
\end{figure}

The current profile can be approximated semiclassically by applying the method of stationary phase to Eq.~(\ref{eq:propagator}). The stationary points are those of the reduced classical action $Et+S_{\text{cl}}({\mathbf r},t;{\mathbf o},0)$, and they represent all classical trajectories of emission energy $E$ that connect the source located at ${\mathbf o}$ with a given destination ${\mathbf r}$ on the detector \cite{TunnelQBM,FeynmanHibbs}. For fixed $E$, all classical trajectories start with initial kinematic momentum $p=\sqrt{2mE}$. If we denote the angle between the $z$-axis and the direction of emission by $\theta$, the classical equation of motion 
\begin{equation}\label{eq:ClassicalMotion}
\rho(t,\theta)=p\sin\theta\,\frac{|\sin(\omega_{L}t)|}{m\omega_L}, \quad
z(t,\theta)   =p\cos\theta\,\frac{t}{m}-\frac{Ft^2}{2m},
\end{equation}
can be used to determine the shape of the envelope of the current profile. According to Eq.~(\ref{eq:ClassicalMotion}), the maximum lateral extension is given by the cyclotron radius $\rho_{\text{max}} = p/(m\omega_L)$ for $\theta=\pi/2$, and $\omega_{L}t=\pi/2$. To obtain the minimum lateral extension we observe that $\rho(t)$ will periodically vanish at $t=T_k$. However, due to the initial momentum, at these instances the trajectories will cover a range of $z$--coordinates from $z(T_k,0)$ to $z(T_k,\pi)$. The number of trajectories connecting the source with a point on the detector is obtained by solving Eq.~(\ref{eq:ClassicalMotion}) for $\theta,t$ with $z,\rho$ fixed. In general, the twofold degeneracy familiar from the purely electric case (Fig.~\ref{setup_exp}) persists, but close to the center of a resonance (given by $z_k = z(T_k,\pi/2)$) there exists a transition region with four classical trajectories, as depicted in Fig.~\ref{res_detail}.  The semiclassical result (left hand side), obtained by summing over classical paths in Eq.~(\ref{eq:propagator}), faithfully reproduces the exact quantum solution available from Eqs.~(\ref{eq:CurrentDensity}) and (\ref{eq:GreenQM}) (shown to the right).  In this figure, the encircled numbers denote the count of classical trajectories in each sector.  They are delineated by two caustic surfaces (solid lines AB and AD) that approximately read in parametrized form $0\le\theta\le\pi$:
\begin{eqnarray*}
\rho(\theta) &\approx& \frac{2E}{F}\left|\frac{\sin^3\theta}{\cos\theta}\right|, \\
z(\theta)    &\approx& z_k -\frac{p}{m}T_k\frac{\cos(2\theta)}{\cos\theta}. 
\end{eqnarray*}
Thus, the constriction is narrowest at $z_k$, where the width $\overline{AC}$ of the focal spot is independent of $\mathbf B$:  $\overline{AC}\approx E/F$.  Interestingly, the elongation of the resonance region $\overline{BD} = (2k\pi p)/(\omega_L m)$ does not involve the electric field $\mathbf E$.  We should point out that the resonances will overlap ($z(T_k,\pi) < z(T_{k+1},0)$) for $F/(pB) < e/(\pi m)$.  In this instance, the number of classical paths may exceed four.

\section{Uncertainty in the focal region} 
\begin{figure}
\onefigure{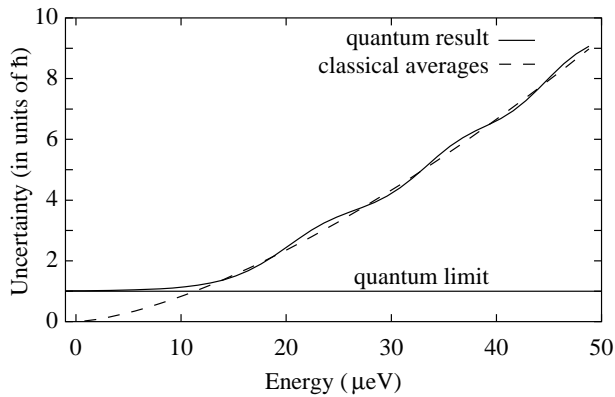}
\caption{Uncertainty $\Delta r_\perp \, \Delta p_\perp$ and classical averages, Eq.~(\protect\ref{eq:average}), as a function of emission energy in units of $\hbar$,
fourth resonance, $F=116$~eV/m, $B=0.001$~T.}
\label{uncertainty}
\end{figure}

At the photodetachment threshold ($E\rightarrow0$), the classical width of the current filament, $\overline{AC} \approx E/F$, approaches zero.  However, the uncertainty principle will put some constraints on the lateral current density distribution.  Let us first elaborate on the classical picture:  Since the emission is isotropic, the average value of $p_{\perp}^2$ is $\langle p_{\perp}^2 \rangle_{\text{av}} = \frac23\,p^2 = \frac43 mE$.  For the mean square width $\langle r_{\perp}^2 \rangle_{\text{av}}$ of the distribution, we integrate over all trajectories $\rho(\theta)$ (Eq.~\ref{eq:ClassicalMotion}) that arrive at $z_k$.  For non-overlapping resonances, this procedure yields approximately:
\begin{equation}\label{eq:average}
\sqrt{\langle r_{\perp}^2 \rangle_{\text{av}}\langle p_{\perp}^2 \rangle_{\text{av}}}
\approx \frac{E}{F}\,\sqrt{\frac{32}{45}\,mE}.
\end{equation}
In a quantal treatment the relevant operators are the canonical momentum operator and the position operator:
\[
p_{\perp}^2=p_x^2+p_y^2,
\quad
r_{\perp}^2=x^2+y^2.
\]
For these operators the relation 
$\sqrt{\langle r_{\perp}^2 \rangle\langle p_{\perp}^2 \rangle}\ge\hbar$
must hold \cite{foot_uncertainty}. In Fig.~\ref{uncertainty} we compare the expectation value $\Delta r_\perp \, \Delta p_\perp=\sqrt{\langle r_{\perp}^2 \rangle \langle p_{\perp}^2 \rangle}$ as calculated from the Green function Eq.~(\ref{eq:GreenQM}) with the corresponding classical average $\sqrt{\langle r_{\perp}^2 \rangle_{\text{av}}\langle p_{\perp}^2 \rangle_{\text{av}}}$ as a function of energy. Once the classical average exceeds the quantum limit $\hbar$, the semiclassical theory provides a good estimate for the uncertainty product (and the current profile in general). For sufficiently small energies $E$, however, a quantal regime prevails: The uncertainty quickly approaches its lower boundary, and consequently, the current distribution becomes Gaussian in shape \cite{Gauss}. Typically, this behaviour becomes prevalent at energies of order $E\approx 1/\beta$; the resulting minimum uncertainty wave function is considerably extended in space, as an estimate of the focal spot radius $\Delta r_\perp \approx 1/(\beta F)$ shows: For the field strengths used experimentally \cite{BlondelI}, a resolution of order $\Delta r_\perp \approx 100$ nm is achieved.

\section{Conclusion} We have studied near-threshold photodetachment in the presence of parallel static electric and magnetic fields. We saw that the magnetic field focuses the photoelectrons into a series of spatially repeating constrictions. The topography of the resonances was calculated and analyzed in terms of classical paths \cite{foot_path}. We demonstrated that the superposition of four classical paths generally accounts for the involved interference pattern present in the constriction sector. However, below a critical emission energy, the current distribution is governed by the uncertainty principle, and the resonances are characterized by a simple Gaussian profile.

\acknowledgments
We appreciate stimulating discussions with C.~Blondel. Partial financial support by the Deutsche Forschungsgemeinschaft is acknowledged.


\begin{thebibliography}{99}
\bibitem{foot_spin}
Strictly, the energy shift $\pm\hbar\omega_L$ caused by the electron spin should be included in Eq.~(\ref{eq:psi_source}). The actual current distribution arises by superposition of both spin components. For weak fields, however, the effects of spin are negligible, and we will omit them here.
\bibitem{foot_AiCi}
Eq.~(\ref{eq:GreenQM}) holds only for $z<0$; otherwise, the arguments of 
$\text{Ai}$ and $\text{Ci}$ must be exchanged.
\bibitem{foot_path}
A quantum path analysis of inelastic processes in time-dependent fields is possible, but less visualizable. See, e.~g., Ref.~\cite{Lewenstein,Kopold,Haenggi}.
\end{thebibliography}
\end{document}